# Analyzing Design Process and Experiments on the AnITA Generic Tutoring System

Matthias R. BRUST
Faculty of Sciences, Technology and Communication, University of Luxembourg
L-1359 Luxembourg, Luxembourg

and

Steffen ROTHKUGEL
Faculty of Sciences, Technology and Communication, University of Luxembourg
L-1359 Luxembourg, Luxembourg

## ABSTRACT

In the field of tutoring systems, investigations have shown that there are many tutoring systems specific to a specific domain that, because of their static architecture, cannot be adapted to other domains. As consequence, often neither methods nor knowledge can be reused. In addition, the knowledge engineer must have programming skills in order to enhance and evaluate the system. One particular challenge is to tackle these problems with the development of a generic tutoring system. AnITA, as a stand-alone application, has been developed and implemented particularly for this purpose. However, in the testing phase, we discovered that this architecture did not fully match the user's intuitive understanding of the use of a learning tool. Therefore, AnITA has been redesigned to exclusively work as a client/server application and renamed to AnITA2. This paper discusses the evolvements made on the AnITA tutoring system, the goal of which is to use generic principles for system re-use in any domain. Two experiments were conducted, and the results are presented in this paper.

## 1. INTRODUCTION

Research on tutoring systems in order to provide methods for more efficient and intense learning in Computer-Based Training (CBT) has attracted a lot of attention. However, investigations have shown that there are many domain-specific tutoring systems that cannot be adapted easily to other domains, because of their static architecture and domain-driven Graphical-User-Interface (GUI). [5] pointed out that traditional tutoring systems are extremely domain-dependant and neither methods nor knowledge can be reused. Evidently, more and more tutoring systems have been developed to match the requirements of different disciplines. As a consequence of this traditional architecture, this work requires several persons to continue maintaining and developing the system, i.e. a knowledge engineer with programming skills.

The general solution for these problems is to reuse the architecture of a tutoring system, and substitute the domain-specific parts with generic components. Our approach has been to first examine existing tutoring systems to extract common functionality. After that, these results have been described in terms of components. This approach, however, also faced problems, because in some cases it has been very difficult to discover and recognize dependencies between functionalities. Sometimes, we have not even been able to classify components because of an extremely interwoven architecture [10]. Based upon these considerations, AnITA, a generic tutoring system, has been developed and implemented. In the generic tutoring system concept proposed, the domain has been separated as far as possible from the architecture of the tutoring system itself.

The first draft of AnITA intended to serve as an environment for both training and testing. We received positive feedback from the use of AnITA, showing that it tackles the problems aforementioned. However, other problems of a different scope have been revealed throughout its use. Therefore, we decided to partly re-design AnITA to deal with these problems as well.

This work reports on the evolvement of the AnITA tutoring system, which aims to use generic principles for architecture re-use in any domain. In section 2, the development and design process of AnITA is described, pointing out new obstacles and problems that appeared. Section 3 introduces AnITA2 as a method to tackle this situation. Experiments were setup and accomplished with AnITA2. Their results are shown in Section 4. Section 5 gives an overview of the new project that integrates AnITA2 with CALM [1] in order to work in a mobile environment based on ad-hoc networks. Section 6 summarizes the experiences and results we obtained during our work with the AnITA tutoring system.

## 2. AnITA: TRAINING AND TESTING

**Concept**
AnITA is intended to serve as environment for both training and testing. In the training mode, students are able to "play" with the questions, see what happens when choosing the wrong answers, and recognizing major misconceptions. In the test mode, students demonstrate their understanding in an "official" test that can be evaluated by a human tutor on a remote computer. To motivate learning, AnITA provides a variety of question types: multiple-choice, calculation-driven, and fill-in-the-blank questions with two different fill-in types (text and combo-box). As already mentioned, the domain has been separated from the architecture of the tutoring system itself. One benefit of such a separation is the possibility to use XML for defining tests and, thus, a step towards establishing well-known standards [8]. Based on such a language, a professor with little or no programming skills is still able to model tests. Furthermore, system-independent domain design will be possible. There will no longer be a need for a knowledge engineer. Usually, tutoring systems are not able to adapt to student's knowledge. It has been a challenge to investigate possibilities of finding generic principles of adaptation. To do that, test paradigms have been examined and realized as described in the subsequent section [4].

**Test-Paradigms**
XML is used as technology to describe test paradigms. AnITA realizes four test paradigms (cf. Figure 2.1). *Free selection* displays questions in the order they appear in the XML file. *Causal Links Selection* is sensitive to right and wrong answers.

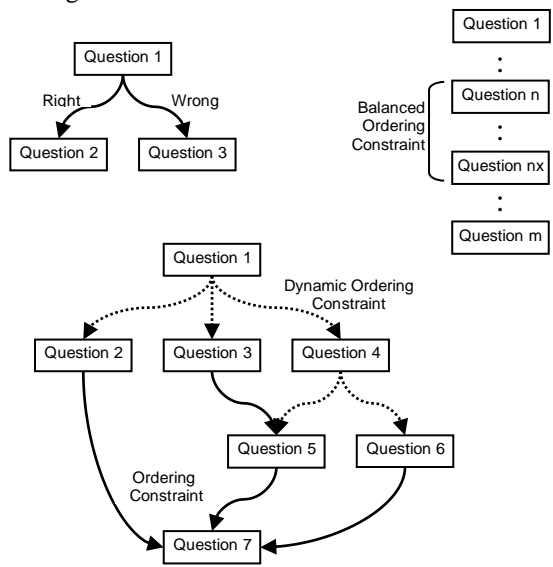

*Figure 2.1: Causal Links, (Dynamic) Ordering Constraints and Balanced Ordering Constraints as Test-Paradigms in AnITA.*

*(Dynamic) Ordering Constrain Selection* follows a pre-determined order for selecting questions. A forced ordering constraint question can only be called upon other question as a reference. This paradigm takes into account the existence of a question that is a subpart of other questions and that according to the context cannot appear alone. A dynamic ordering constraint selection enables the system to choose randomly from different constraints. This feature is aimed to create different tests out of just one XML-file. *Balanced Constraint Selection* is more basic from a conceptual point of view. The attribute balance is introduced with its values **n** and **p**. The **n** value implies an arithmetic average **a** for the last **n** selected questions. If **a** is greater than **p** the system may continue selecting the next question and following the ordering constraints, causal links or free selection. Otherwise, the selection repeats the last **n** questions.

**Implementation and Architecture**
AnITA is a pure Java stand-alone application based on the Swing GUI that requires the Java Runtime Environment to run. With this application, students can apply their knowledge in the available questionnaires. Interactions are evaluated, but neither stored on the system's internal state nor on a local database. In the testing mode AnITA can establish a connection to a remote database through a Servlet that runs on the remote host. Because the Servlet is written in Java and the implemented database has a rational design, a JDBC/ODBC-driver is used to set up the communication between the Servlet and the database. The tests are stored on the client (training mode) or on the server (testing mode) as illustrated in figure 2.2.

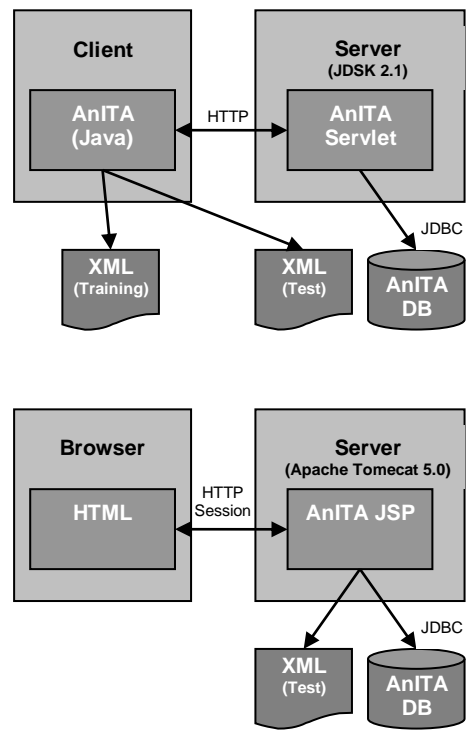

*Figure 2.2: Overview of the AnITA architecture (left) and AnITA2 architecture (right)*

**Experiences**
As reported in [3], first experiments with AnITA have shown that students expressed mostly positive comments. At this time, we were able to apply four domains in AnITA: Operating Systems, Statistics, Architecture, and Information Systems. We understood this as indication that we were on the promising way on creating a generic tutoring system successfully. Although these experiments were seen as encouraging for AnITA, nowadays, we have to re-evaluate AnITA from a wider perspective and concede that important aspects had not been evaluated.

To install AnITA on the university's computer, we needed a computer expert who knows how Java programs work and how to install the Java Runtime Environment. AnITA had to be installed on all 28 computers in the lab. Although these steps may appear very simple, we must not forget to design applications as simple as possible – including their installation procedures.

Furthermore, because of the overall design (Swing, XML parser, complex data structures etc.), AnITA runs very slowly on older computers that can be found quite often in public labs. It took between fifteen seconds and one minute to launch AnITA on these computers, obviously an undesirable situation.

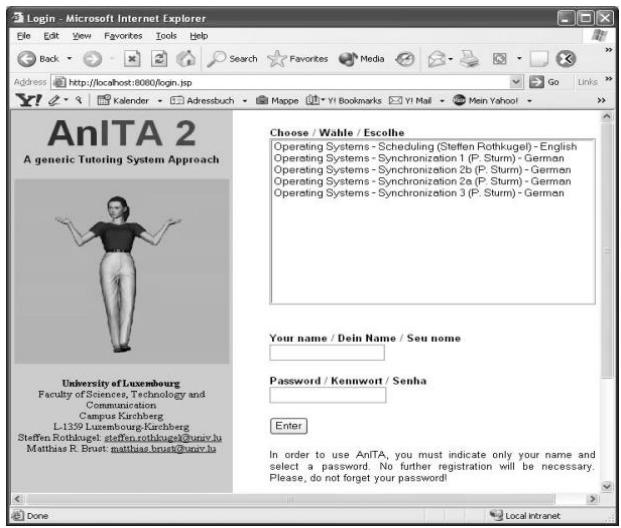

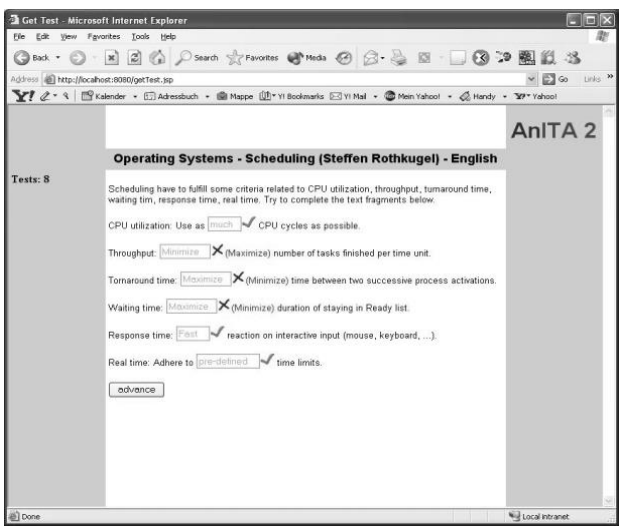

*Figure 3.1: Login page and a fill-in question with its correction*

Finally, another design flaw became apparent that we found unsatisfactory from a pedagogical point of view: To evaluate the system and the student, it would be best to have as much information as possible about both [8]. However, in AnITA, the student could act autonomously in the training mode without ever publishing his/her results and behavior. The pedagogical value of AnITA decreased because it was almost impossible to receive data about the student's performance since the last tests. Even though it was satisfying having reached our goal, it was disappointing to recognize that AnITA caused such serious problems. As we observed these problems, new technologies became available and we decided to re-design AnITA. As a result, AnITA2 has been implemented.

### 3. AnITA2: EVALUATION-BASED DESIGN

**Concept**
The basic concept of AnITA did not change, but it was extended in some areas. As described in the previous section, AnITA provides a variety of question types.

One concrete additional demand has been identified due to the following case: a physician analyzed AnITA and expressed the desire to interact with x-ray pictures to locate illness areas. In AnITA2, we took this into account by designing the system in a component-based way with respect to the question types. Now, it is easy to create new questions types in AnITA2, because it is not necessary to change the application code. Additionally, a performance meter was added. This enables students to compare visually his/her performance in a bar chart that shows the performance of the last 20 training units of one specific test (Figure 3.1).

**Implementation & Architecture**
AnITA2 is designed on client/server principles as a Web-application using Java Server Pages (JSP). Therefore, AnITA2 does not have to be installed, because it can be accessed using a standard Internet browser. Tests are exclusively stored on the server-side and are transmitted to the client on demand (on a question-by-question basis). All data is stored in a rational database on server side. AnITA2 uses component-based techniques for extensibility with respect to question types. Internally, introspection is used to implement this feature. Introspection allows Java code to discover information about the fields, methods and constructors of arbitrary classes and to dynamically invoke them.

```
public String setTest() {
    // Choose next question
    ...
    // Initialize the class loader
    MultiClassLoader loader = null;
    loader =
        new FileClassLoader(GlobalValues.CLASSES_PATH + File.separator);
    ...
    testClass = loader.loadClass("anita/"+n.getNodeName());
    ...
    testObject = testClass.newInstance();
    ...
    String testInHTML = "";
    ...
    Class[] classPara = new Class[1];
    classPara[0] = org.w3c.dom.Node.class;
    Method testMethod = testObject.getClass().
                        getMethod ("set"+n.getNodeName(), classPara);
    testInHTML += (String) testMethod.invoke(testObject, new Object[] { n });
    ...
    return testInHTML;
}
```

*Figure 3.2: AnITA2 code that invokes component-based question types.*

While reading the XML-based questions, AnITA2 needs to execute the appropriate code related to that question type. Questions types are not hard coded directly in the AnITA2 application. Rather, question types can be added dynamically by simply changing the DTD specification and adding a new component to the system. Figure 3.2 shows the code that invokes the question types as described above.

### 4. EXPERIENCES WITH AnITA2

Since AnITA2 is a generic approach, it is important to validate the results with a variety of experiments. In order to do this, we set up an experiment in a language school as soon as AnITA2 became stable. Here, AnITA2 provided basic questions to teach German to Brazilians. AnITA2 was improved by using the results of this experiment and another one was set up that focused on more in-depth questions regarding the course "Operating Systems" at a European university.

**Language Test**

One experiment was done in a language school in Brazil. All nine students were Brazilians and between 18 and 24 years old. They had attended their initial German course for eight weeks having two hours of lectures per week. All students were well experienced with computers. The questions were simple and covered a wide range.

The objective of our test was to find out, if students who paid to learn English would use a free online-learning tool. We also prepared AnITA2 to find answers to the following questions.

- Are there indicators that students would use this free tool on their own?
- In which way would they use AnITA2? (When? How long? How many times? Progress?)
- Which influence does the use of AnITA2 have in the classroom?

They could take the test by using all of the teaching material and could use as much time they wanted. We did not explain that it was a test to show if they really would use AnITA2 and we did not mention that any data was stored in a database.

*Result:* Most of the students (six students) discontinued the test between the eighth and thirteenth question.

*Conclusion:* We wonder, if students prefer a certain number of questions for being more motivated than with a different number and in which way the time they spend on the exercises is influencing on this number. Next experiments have to show this correlation.

*Result:* Two students commented that it was problematic not to know how many questions remained.

*Conclusion:* The DTD-Specification of AnITA2 and AnITA2-code were modified to switch on the option to show the remaining number of questions on the left side of the AnITA page. Modifying the specification, it has to be observed that the number of remaining question changes dynamically when using *Causal Links*, *(Dynamic) Ordering Constraints* and *Balanced Ordering Constraints*. For this, it will be necessary to readapt the new feature to realize test paradigms.

*Result:* There was also the desire to present all question on one page.

*Conclusion:* Here we recognized the desire to change AnITA2's format to a traditional questionnaire. Furthermore, after we asked more specific questions, we discovered that the students hoped to find the answer to a question by reading other questions (hints, etc.). But we argued that this situation would complicate the creation of questionnaires because one question cannot be used to answer another question and this would restrict AnITA2.

*Final Conclusion:* Most students had a job aside from doing their studies. Hence, they sometimes could not attend the classes. These students in particular used AnITA2 in their free time to try to make up for what they missed. Therefore, there is a need for an extra option such as this. During the lectures, students started discussing about the AnITA2 system. This seemed to encourage other students that did not use AnITA2 before to start using it, probably because they wanted to be able to follow the discussion. Comparing the test results collected, we discovered that a lot of students kept on using AnITA2 until they had answered all the questions correctly. Therefore, we conclude that AnITA2 can be seen as a method to encourage learning.

**In-depth Test for Operating Systems**

On the basis of the modifications, which resulted from the first experiment, we conducted a second experiment. Our domain was the discipline of synchronization as subpart of operating systems in an in-depth level. The participants were three students between the ages of 24 and 28 that have a few days to their final diploma exams. The main questions were:

- Does AnITA2 have the capacity to design an in-depth test?
- Would the results of the test match the professor's evaluation for the same student?
- What is the student's opinion about using AnITA2 for testing or training?

They could not use the teaching material, but could take as much time as they wanted. We did not explain that it was a test to show if they really would use AnITA2 and we did not mention that any data would be stored in a database.

*Result:* All students confirmed that testing their knowledge on AnITA2 was very satisfying, because they increased their self-confidence for the "real" test. They mentioned that they prefer AnITA2 to other multiple-choice-based tutoring systems, because of the variety of question types. Interestingly, one of them started explaining that he felt more difficulties with pure text inputs than with multiple-choice questions. Other students agreed with his opinion on this issue.

*Conclusion:* Our decision to offer a component-based system for question types is supported by some students' opinions. The classification of question types we did before the design of AnITA was understood intuitively by the students [2].

*Result:* The desire was expressed that the tool could adapt to the student's knowledge.

*Conclusion:* In intelligent tutoring systems, adaptation is an important role, but we see it as a challenge to continue working on our generic adaptation methods, although we focus on a tutoring system.

*Result:* One student pointed out that AnITA2 is surprisingly capable to testing algorithms and fragments of code in an interesting way, but he also mentioned that these questions took a lot of time.

*Conclusion:* We conclude that it was unavoidable to add specific functionality to a generic tutoring concept in order to widen the scope. This could also be a hint of the limits of the generic concept used in AnITA and AnITA2.

*Final Conclusion:* It is difficult to compare the student's performance on AnITA2 with the professor's evaluation, because the system offers an in-depth written test while the final diploma exam is an oral test that covers the entire discipline. Comparing AnITA's evaluation to the professor's, the student tended to receive a higher grade from the professor. We were not sure about the significance of this result, so we decided to prepare another experiment to investigate this point more thoroughly. Finally, the graphical performance meter was seen as very helpful for self-evaluation.

## 5. FUTURE WORK

More and more students use palmtop and handheld computers at home and at university. With larger memory capacities, a variety of data input devices, and the ability to link into wireless networks, applications from different domains must be adapted to run in mobile environments.

In the future, we will work on this challenge by developing CARLA, a learning system for mobile ad-hoc networks. In contrast to the AnITA system, the purpose of CARLA is to support cooperative learning for two reasons. First, because there are extreme constraints on information dissemination in ad-hoc networks [6], a cooperative concept has a better chance of being successfully implemented. On the other side, we see this restriction as an opportunity to gain experience with post-modern learning theories, i.e. cooperative and collaborative learning in a computer-based mobile environment.

**Scenario**
For example, students might join forces to prepare for exams. Teaching materials like lecture notes, slides, and basic questions, are distributed to students at specific locations like the lecture room or professor's office. This is done by a professor's mobile device or by a stationary node. Initially, all students start using the same teaching material. During a lecture, students can write annotations on the slides [1]. Later they may want to test their knowledge by answering questions. Through this process, they may discover a correlation between sections of the teaching material and their annotations and questions. They can express their findings by adding links. The resulting personalized material enables students to gain a deeper understanding of the subject.

As a cooperative environment, the CARLA system aims to disseminate all additional material to the students. CARLA enables students to analyze received material by using evaluation mechanisms. The resulting evaluations will be shown on annotations, links, questions and teaching material, to represent the "usefulness" of an element. Students are able to recognize misleading or falsified content more easily. Figure 5.1 gives an example of teaching material combined with additional elements.

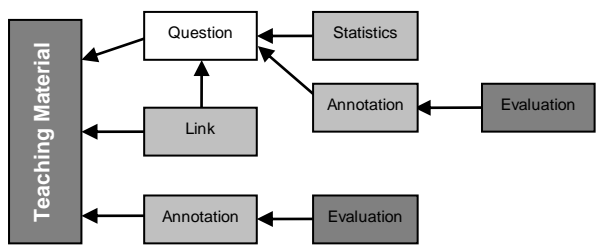

*Figure 5.1: Illustrating the correlation between questions, annotations, links and evaluations*

CARLA is designed to run with short periods of interaction between the devices in the ad-hoc network. This design is important in order to take into account the restrictions of a mobile ad-hoc network.

The objective of CARLA is not only for professors to be able to evaluate students' cooperative work, but also to steer the learning process in the right direction with additional links and/or annotations [7]. Another benefit of CARLA is that the professor will be able to redesign the initial teaching material based on the students' links, questions, and annotations, thereby increasing the scope and usefulness of the teaching material.

## 6. CONCLUSION

In this paper, the design and evolvement of the AnITA generic tutoring system is described. The concept proposed intentionally separates the domain from the tutoring system. The main benefit of this system is that it creates the possibility to apply an XML-based language for tests and to establish standard directives for system design [8]. Based on the resulting language, even professors with little or no programming skills are enabled to model tests, and both the domain and tutoring system can be designed and maintained independently.

Experiments with AnITA showed that it was necessary to re-evaluate AnITA from a broader perspective, and concede that important aspects had not been considered in both the evaluation and the design process.

While adapting AnITA to AnITA2, the code became about 25 % shorter and more concise. We concluded that this was the result of using a Web browser as user interface. The browser realizes the GUI through based upon HTML code that is effectively plain text.

Several years ago, researchers were faced with the challenge of realizing learning environments as an Internet-based client/server application. Since the use of mobile systems is increasing, the focus has shifted to the development of self-organized learning platforms for such systems. CARLA was proposed as a way to deal with this new aspect. We hope that CARLA can continue the work that the AnITA generic tutoring system has begun. Finally, the results of our experiments indicate that AnITA2 seems to be a catalyst for increasing the efficiency and intensity of the learning process.

## ACKNOWLEDGMENTS

This research is supported in parts by the Luxembourg Ministère de la Culture, de l'Enseignement Supérieur et de la Recherche.